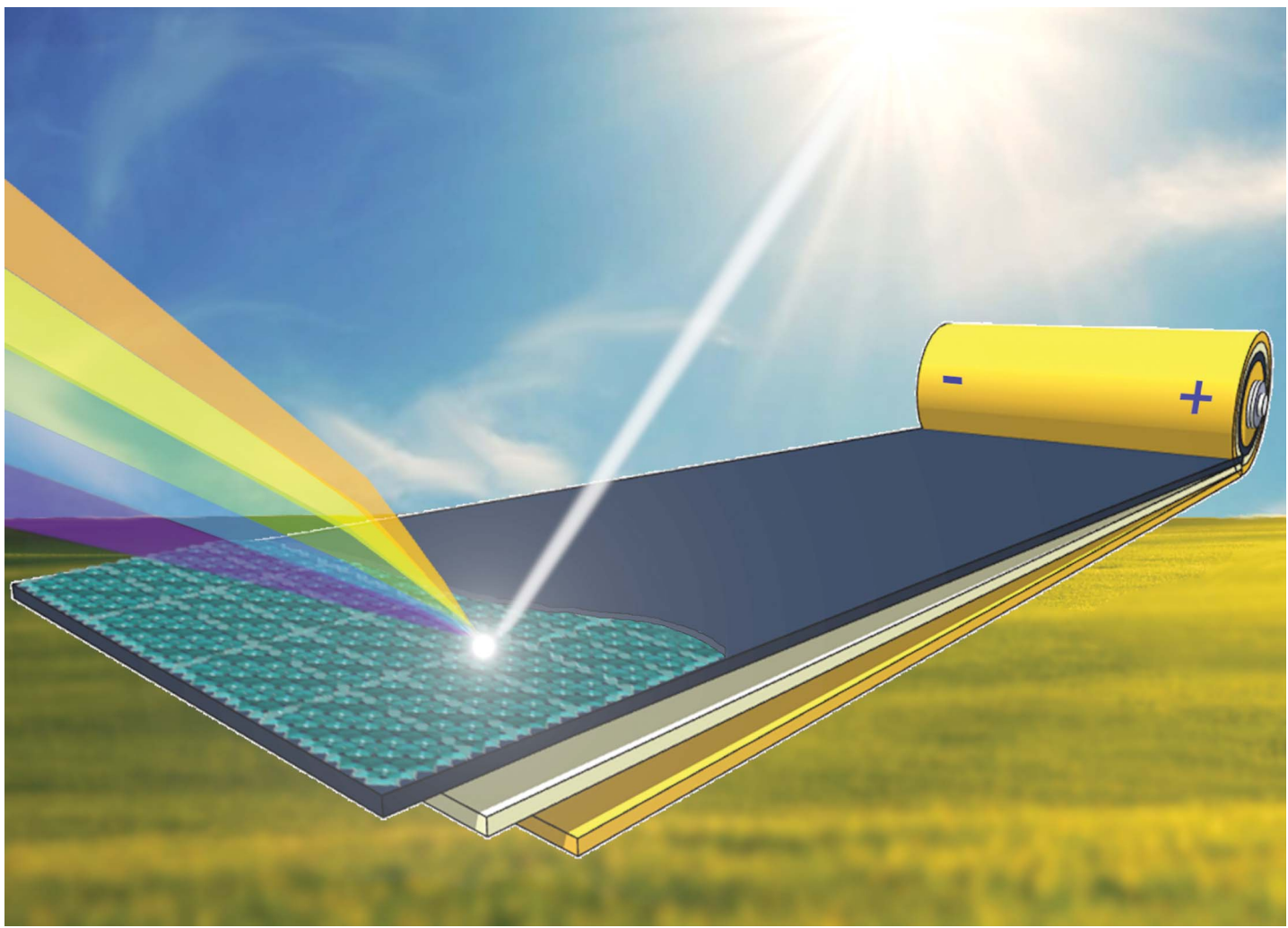

Showcasing research from Professor Tarancón's laboratory, Catalonia Institute for Energy Research (IREC), Catalonia, Spain.

Operando probing of Li-insertion into $LiMn_2O_4$ cathodes by spectroscopic ellipsometry

The work presents a novel Operando Spectroscopic Ellipsometry tool for the characterization of battery materials combining high spatial resolution with multi-layer and time-resolved capabilities. A proof of concept is shown where the lithium content of a $LiMn_2O_4$ thin film is tracked along voltage sweeps and fast potential steps.

### As featured in:

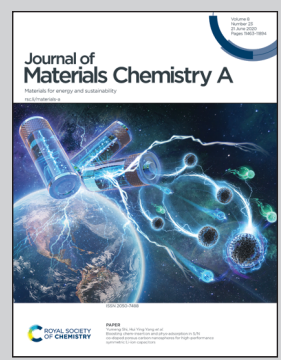

See A. Morata *et al.*,
*J. Mater. Chem. A*, 2020, **8**, 11538.

ROYAL SOCIETY OF CHEMISTRY

rsc.li/materials-a

Registered charity number: 207890

# Operando probing of Li-insertion into LiMn$_2$O$_4$ cathodes by spectroscopic ellipsometry†

A. Morata, [*a] V. Siller,[a] F. Chiabrera,[a] M. Nuñez,[a] R. Trocoli,[a] M. Stchakovsky[b] and A. Tarancón [ac]

A novel non-destructive methodology for *operando* observation of ion intercalation and the state of charge on battery electrodes is presented based on spectroscopic ellipsometry (SE). The potentiality of this technique for performing time-resolved measurements of (de-)lithiation processes on electrode materials has been demonstrated using thin film spinel LiMn$_2$O$_4$ as a cathode for Li-ion batteries. The chemical diffusivity of Li$^+$ ions in this material has been determined by the time evolution of the Li insertion into the film, and in the light of these results, it has been possible to relate the controversial pseudo-capacitive behavior observed in this material to the nanostructure of the layer. The results presented here establish a new way for *operando* characterization of battery materials and devices providing a powerful tool for the understanding of ion diffusion mechanisms in a collection of electrochemical devices.

## Introduction

Batteries are currently a mainstream high energy storage technology ubiquitously adopted in diverse applications ranging from electronic devices to vehicles and domestic energy storage. These new energy demands foster continuous advances for improving devices by achieving higher capacity, faster charge–discharge capabilities and longer lifetime. Advanced concepts in batteries typically involve exploring novel materials and interfaces and the related electrochemical and diffusive processes taking place during the operation of the device. In particular, it is crucial to gain insights into dominant interface phenomena such as Li$^+$ insertion, formation of solid electrolyte interphases (SEIs)[1,2] or ion blocking grain boundaries.[3–8] However, the continuous evolution of interfaces during the operation of devices makes it difficult to obtain reliable information only from post-mortem analyses after multiple cycles. For this reason, a lot of effort has been dedicated in the last decade for developing novel techniques to obtain new insights into interface phenomena *in operando* and therefore, into their direct correlation with electrochemical performances.[9–13] Despite these efforts, some of the most powerful techniques, such as isotopic ion exchange methods,[14] *in situ* TEM[14] and the collection of synchrotron radiation based techniques,[10,11,15–20] are highly sophisticated limiting the straightforward access to essential information for developing highly performing batteries. In addition, a number of commonly available techniques have also been explored including X-ray diffraction,[21–23] atomic force microscopy (AFM),[5,24] Raman spectroscopy[25–27] and Fourier transform infrared (FTIR) spectroscopy[28,29] showing different advantages and limitations regarding spatial and time resolutions. On the other hand, despite the well-known capabilities of Spectroscopic Ellipsometry (SE) for the study of the properties of thin film systems (including multi-layered devices), the use of this affordable technique in the field of Li-i is very limited[30] and, to the best of the authors' knowledge, it has never been implemented for real *in operando* measurements in the battery domain. SE measures the change in the polarization state of a light beam reflected in the sample and compares it to a model. In this way, relevant optical properties of the material such as the dielectric function ($\varepsilon$) can be obtained as a function of the wavelength.[31] This technique is typically used to determine the properties of thin films and also complex multilayer systems, being sensitive to parameters such as crystallinity, materials ratio in mixtures, roughness, structure of the interfaces, *etc.*

In this work, an optical non-destructive methodology based on SE is proposed for the first time to provide information on the (de-)lithiation phenomenon occurring in a model electrode such as LiMn$_2$O$_4$ operating as a cathode in a Li-ion battery. Spinel LiMn$_2$O$_4$ is a low cost/low toxicity compound based on abundant raw materials showing high working potential and safety together with a competitive theoretical capacity (148 mA h g$^{-1}$).[32,33] Furthermore, this material has shown an outstandingly fast performance when presented in the nanostructured form, the origin of which is still a matter of controversy.[34–41] This feature has important practical implications as it would allow development of devices that can simultaneously provide high power typical of capacitors, and high storage capacity expected from batteries.[42] Due to the fast response of the SE system and the strong effect of the oxidation state on the optical properties of compounds, SE has been used in this work to monitor Li$^+$ transport properties by real-time tracking of the valence changes associated with lithium insertion–deinsertion along LiMn$_2$O$_4$ thin films and interfaces. Furthermore, SE is used to accurately measure the thickness and porosity of the films giving rise to values of effective diffusion of Li cations in LiMn$_2$O$_4$.

[a]*IREC, Jardins de les Dones de Negre 1, Planta 2, 08930, Sant Adrià del Besòs, Spain. E-mail: amorata@irec.cat*
[b]*HORIBA Scientific, Avenue de la Vauve, Passage Jobin Yvon, 91120 Palaiseau, France*
[c]*ICREA, Passeig Lluís Companys 23, 08010, Barcelona, Spain*

† Electronic supplementary information (ESI) available: It includes structural analysis, details on the ellipsometric model construction and fitting procedure, triggering acquisition method and FEM simulation details. See DOI: 10.1039/c9ta12723b

## Results and discussion

LiMn$_2$O$_4$ thin films were grown by pulsed laser deposition (PLD) on 80 nm-thick sputtered Pt layers deposited on top of 10 × 10 mm Si chips. A multilayer approach has been used to compensate the Li loss typically occurring in PLD deposition of lithium compounds to obtain the stoichiometric spinel phase (details on this procedure can be found elsewhere[43,44]). Structural and morphological characterization of the multilayer structure and the LMO thin film has been carried out by SEM and AFM and the results are presented in Fig. 1. The SEM images show relatively large crystalline domains and a thickness of 330 ± 10 nm for the LMO layer. According to X-ray diffraction measurements presented in Fig. S1,† the layer shows the appropriate spinel phase without secondary phases.

The as-prepared samples have been brought into contact with Al foil protected with a Kapton® film and introduced in a home-made electrochemical measurement chamber, provided with optical windows and filled with a 0.2 M Li$_2$SO$_4$ liquid electrolyte (a detailed description of the setup can be found in the Experimental section and elsewhere[42]). Cyclic voltammetry (CV) measurements have been conducted, showing very well defined (de)lithiation double peaks typical of the spinel structure (see Fig. 2a). The results are consistent with profiles previously reported for analogous samples, which demonstrates high capacity retention and outstandingly fast performance typically attributed to the so-called pseudocapacitive phenomenon.[42] Simultaneous with the electrochemical characterization, spectroscopic ellipsometry measurements have been carried out between $U$ = 0.350 and 1.05 V, corresponding to completely lithiated and de-lithiated states, respectively (Fig. 2b). The important variations observed in the spectra are mainly due to the change in the Mn oxidation state in LMO, from Mn$^{4+}$ to Mn$^{3+}$, occurring during the Li$^+$ extraction. The significant change observed in the spectra indicates the high sensitivity of SE in monitoring the Li$^+$ insertion/deinsertion process. In order to describe the sample in any oxidation state within the proposed voltage range, simple but flexible geometrical and physical models have been defined. The geometrical model consists of a dense film on top of a flat and completely opaque Pt substrate (see the sketch in ESI S2† and the inset in Fig. 2c). The porosity of the film has been modelled by using a partially empty top layer filled with the liquid electrolyte present in the electrochemical cell, whose optical properties have been measured using a *prisma* setup (as detailed elsewhere[45]). On the other hand, the physical model uses Tauc-Lorentz parameterization of the optical constants of the $LiMn_2O_4$ with four oscillators.[46] A four-component configuration is the simplest model found that is able to track all the changes of the film during cycling. Previous knowledge on the spinel $LiMn_2O_4$ and $MnO_2$ band structures (*i.e.* the material in lithiated and delithiated states) has simplified the selection of the initial optical parameters giving them a clear physical meaning.[47] In particular, the footprint for following the $Li^+$ insertion/deinsertion phenomena is expected to be linked to the adsorption bands at 2.8–3 eV and 3.4–3.6 eV,[48,49] corresponding to the electronic transition from $O^{2-}(2p)$ to $Mn^{4+}(e_g)$ and $Mn^{3+}(e_g)$, respectively (see details on the oscillator model and fitting procedure in the Experimental section and ESI†). The experimental data and the fitting of the proposed model are in great agreement with the range of applied voltages of $U = 0.350$–$1.05$ V, therefore allowing the calculation of the evolution of the optical constants of LMO during cycling of batteries (Fig. 2b and S2c†).

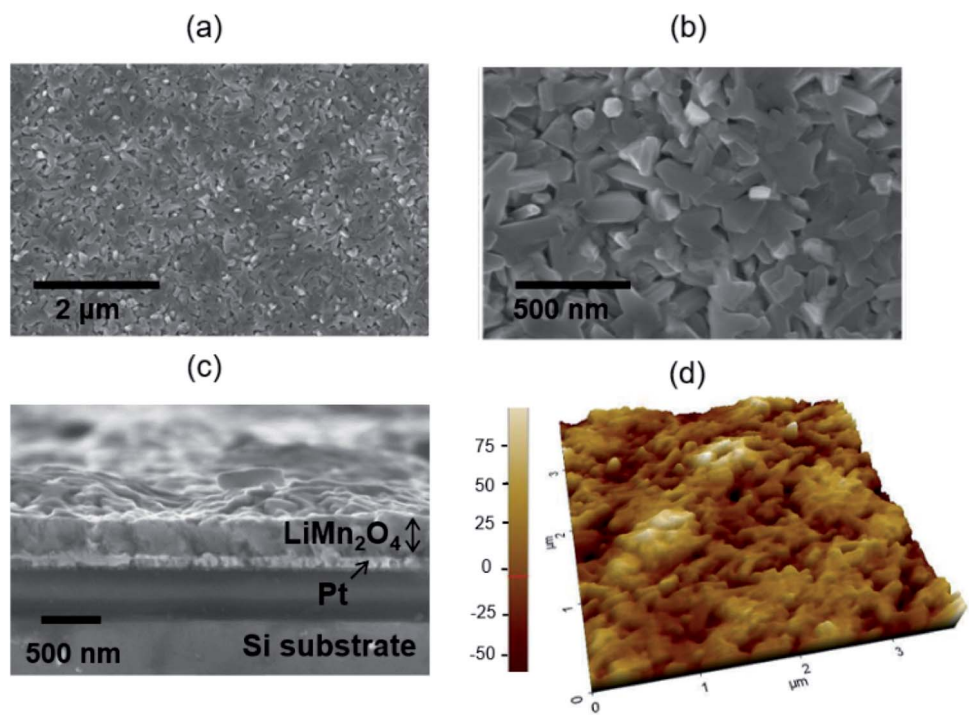

Fig. 1 Microstructural characterization of the LMO deposited film: top-view (a and b) and cross section (c) SEM images of the PLD deposited LMO thin films on top of Si/Pt substrates. (d) TEM cross section of a lamella of LMO/Pt/Si half-cells.

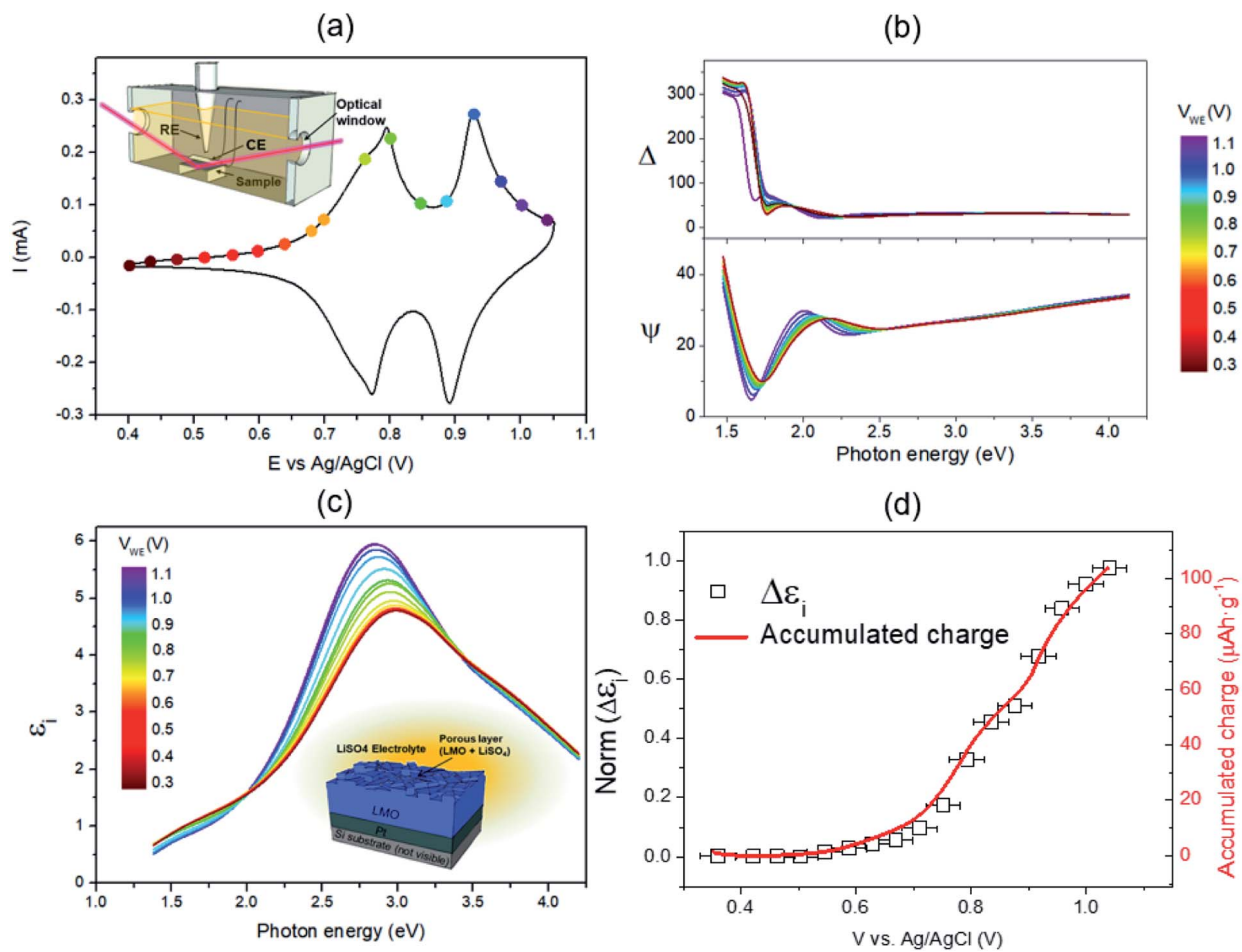

Fig. 2 Combined electrochemical and ellipsometry characterization of LMO films: (a) cyclic voltammetry of a film measured in the specially designed electrochemical/optical *operando* test chamber (see insert). The dots represent the voltages at which ellipsometry spectra were acquired. (b) Evolution of the *operando* spectroscopic ellipsometry signal (Δ and ψ) during the voltammetry (the voltage of the cell during the acquisition is represented in the colour scale bar). (c) Imaginary part of the dielectric constant obtained from the fitting of the obtained spectra obtained during the voltammetric cycle (the voltage of the cell during the acquisition is represented in the colour scale bar). A sketch of the geometrical model employed is presented in the inset. (d) Evolution of $\varepsilon_i$ and the specific charge accumulated in the LMO layer (capacity) during the voltammetric cycle.

Fig. 2b and c show the evolution of the SE spectra with time and the calculated optical constant during the galvanostatic charging of the battery (only the imaginary part of the dielectric function, $\varepsilon_i$, is represented for the sake of simplicity). The optical properties of the multilayer structure are not significantly evolving out of the insertion potentials, namely, $V = 0.75$ and $0.95$ V, while, at these potentials, the light absorption around 3 eV is clearly enhanced. This undoubtedly indicates that the oxidation/reduction of the film is directly correlated with its optical properties, which is fully confirmed by the identical change of the dielectric constant and the charge accumulated in the film during the voltage sweep (essentially proportional to $Li^+$ extraction), see Fig. 2d. The coincidence of both magnitudes allows the direct quantification of the Li content of the electrode layer at any time and applied voltage, *i.e.* the state of charge of the battery, by simply integrating the absorption band at 2.8 eV of the dielectric constant as obtained by non-destructive SE.

Complementary to the battery capacity and state of charge analysis, relevant mass transport information such as diffusivity was obtained for LMO by SE through the analysis of the transients between different charge states. Using an innovative triggering strategy to reconstruct fast transient sign along battery cycling (see the Experimental section/ESI† for further details), sample rates as high as 10 Hz (1 sample every 100 ms) were possible for relatively rough and highly absorbing LiMn$_2$O$_4$ layers. However, it is important to notice that there is no fundamental limitation that prevents reduction in this rate above 100 Hz in a conventional setup, as long as better conditions are provided (the sampling frequency depends on the signal-to-noise ratio, which is governed by experimental features of the sample and the system such as absorbance or surface roughness and electrolyte absorption or optical windows, respectively) or even increase in this frequency using advanced multi-wavelength ellipsometers (able to acquire the whole spectra at a time).

An experiment to induce complete de-lithiation of the LiMn$_2$O$_4$ layer was carried out by applying a potentiostatic step (with voltage changing from 0.35 to 1.05 V vs. Ag/AgCl). The corresponding transient of the dielectric function calculated after SE measurements is shown in Fig. 3a. A sudden change in the dielectric function at the moment of applying the voltage step is clearly observed. This observation can be quantified by integrating the optical absorption band at 2.8 eV, which yields the amount of Li ions inserted into the layer at every single point of time (Fig. 3b). From the obtained curve, an ultrafast oxidation process of LMO is derived showing an insertion of more than 80% of the final charge in less than 2 seconds. This result is consistent with the fast electrochemical performance previously observed for the same thin film cathodes by our group[42–44] and for other similar cathodes in the literature.[34–41] It has been speculated that shorter diffusion paths[39,42] and/or intimate contact with the current collector[41] are behind the fast kinetics leading to pseudocapacitive behaviour.

Here we unequivocally demonstrate this fact, as the accumulated charge (represented by the green line in the plot of Fig. 3b) remarkably coincides with the variation of $\varepsilon_i$ at 2.8 eV, which indicates that all of the introduced charge is immediately invested in the electrochemical oxidation of the bulk LMO (Mn$^{3+}$ to Mn$^{4+}$) without the need for any additional surface accumulation capacitive phenomena.

To further corroborate this point and to quantify mass transport properties of this material, the Li insertion as a function of time obtained by SE has been adjusted to a classical diffusion model calculated by the finite element method (see the ESI† section for details on the procedure). A simple geometry consisting of a cylinder representing an average columnar grain of the film on top of a fully blocking substrate has been used. The obtained results are represented by a green line in Fig. 3b showing excellent agreement with the experimental data therefore confirming the diffusional origin of the observed phenomena. According to the fitting, a chemical diffusivity for Li$^+$ into LMO of $D_{chem}$ 3.5 × 10$^{-12}$ cm$^2$ s$^{-1}$ is obtained, which is comfortably positioned within the wide range of values found in the literature for this particular material (ranging from 10$^{-12}$ to 10$^{-10}$ cm$^2$ s$^{-1}$ when measured with CV, potentiostatic intermittent titration or impedance spectroscopy methods).[3,50–57] All in all, from the here presented results, it is demonstrated that apparently ultrafast Li$^+$ deinsertion, potentially attributed to the pseudo-capacitive effect, can indeed be explained by using a special microstructure, providing a high surface area for Li$^+$ interchange with the electrolyte, even considering a modest diffusivity value. This first example highlights the potential of spectroscopic ellipsometry as a unique technique to carry out fundamental and practical studies of materials for batteries and battery devices *in operando* even when ultrafast phenomena are taking place.

## Experimental

The deposition of LMO thin films was carried out using a Large Area PLD5000 from PVD Products, Inc. equipped with a KrF excimer laser with 248 nm wavelength. Multilayer deposition was used to compensate the Li loss, using LiMn$_2$O$_4$ and Li$_2$O

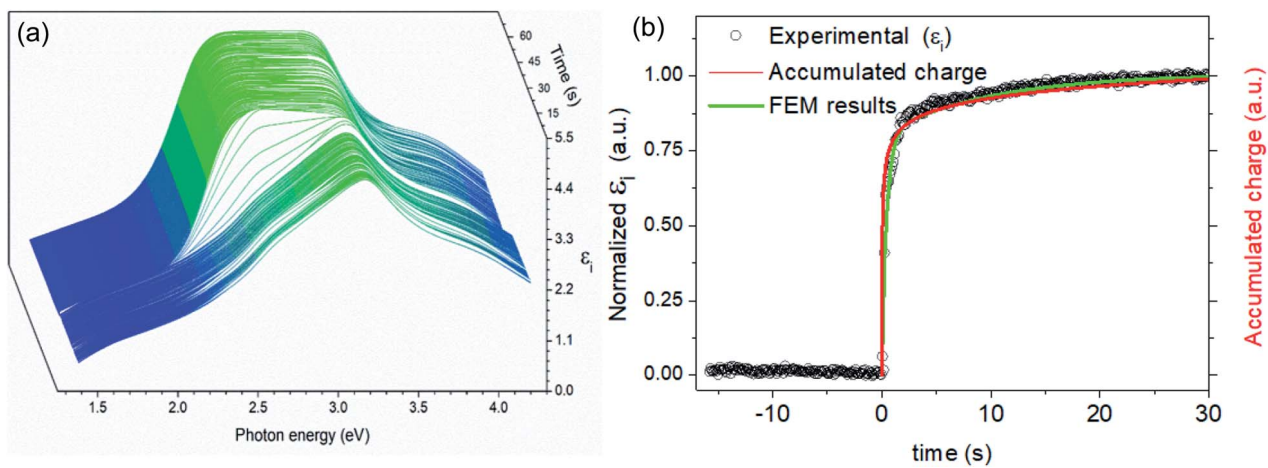

Fig. 3 Transient analysis of the LMO during cycling: (a) variation of the $\varepsilon_i$ of the LMO thin film during a sudden potential step from 0.35 to 1.05 V vs. Ag/AgCl. (b) Evolution of optical absorption at 2.8 eV of the dielectric function during this volt.

<vertical_text>Open Access Article. Published on 05 June 2020. Downloaded on 5/27/2021 4:33:26 PM.
This article is licensed under a Creative Commons Attribution-NonCommercial 3.0 Unported Licence.</vertical_text>

targets purchased from Neyco, France. Details about this procedure have been published elsewhere.[43] Films were deposited onto Si/TiN (10 nm)/Pt (80 nm). The substrates were subjected to a cleaning process before deposition consisting of a subsequent batch in acetone, isopropanol and deionized water. The PLD process was performed in an oxygen atmosphere at a pressure of 20 mTorr. The deposition temperature was 650 °C. A laser fluency of 650 mJ cm$^{-2}$ with a pulse frequency of 10 Hz was used. LiMn$_2$O$_4$ and Li$_2$O materials were deposited alternately in a pulse ratio of 2 : 1 until the desired thickness was reached. XRD measurements were carried out using Bruker-D8 Advance equipment with a Cu K$\alpha$ radiation source and a Lynx Eye detector. SEM images were taken using a Zeiss Auriga. Spectroscopic ellipsometry measurements were recorded by using a phase modulation system (UVISEL equipped with a Horiba spectrometer) at wavelengths ranging from 260 to 2100 nm (4.77 to 0.59 eV). The ellipsometer beam angle of the incidence was fixed at 70.0°, and the spot size was fixed at 2 mm$^2$. A homemade electrochemical chamber was fabricated with a Prusa i3 style 3D printer, using Acrilonitril Butadiene Styrene. The frame of the chip (having the Pt layer exposed) was brought into contact with Al tape. The whole sample is covered with insulating Kapton tape, except for an 8 mm$^2$ slide in the center, which will be the active region in contact with both the electrolyte and the ellipsometer light. The chamber has two optical windows perpendicular to the light beam, made of CaF$_2$ from Thorlabs, providing a transmission window from 18 nm to 8 μm. In order to demonstrate that the chamber does not interfere with the results, a reference sample consisting of a bilayer of thermal oxide (*ca.* 100 nm) and silicon nitride (300 nm) on a Si (100) chip is consecutively measured both inside and outside the chamber. The results are identical, showing differences of measured thicknesses below 1 nm. Moreover, a reference measurement is always taken on the sample under test out of the chamber before the *operando* experiments. These results are compared to the ones obtained inside the cell, discarding any unimportant discrepancy. The chamber is covered with a liquid electrolyte (0.2 M Li$_2$SO$_4$) at a level above the optical windows. Counter and reference electrodes are introduced (Pt mesh and Ag/AgCl, 3 M KCl, respectively).

The effect of nanoporosity on the lithiation mechanism of LiMn$_2$O$_4$ thin films was investigated by FEM analysis. The simulations were performed by using COMSOL Multiphysics using the Transport of Diluted Species module. Details about the procedure are presented in ESI† section 4.

## Conclusions

Spectroscopic ellipsometry has been presented as a new powerful tool for conducting *operando* measurements of Li-ion battery materials and devices. In particular, the evolution of the optical properties of LiMn$_2$O$_4$ cathodes has been obtained during cycling in a liquid electrolyte electrochemical cell. On this basis, the optical absorption of the film has been directly correlated with the Li insertion/deinsertion in steady states of battery charge, enabling a novel non-destructive measurement of the remaining capacity. Furthermore, the technique has proved its suitability for tracking the Li$^+$ incorporation into a LiMn$_2$O$_4$ film during a fast discharging transient of only two seconds. This transient behaviour has been quantitatively described by using a FEM simulation, which allowed the determination of the effective diffusion coefficient of Li$^+$ cations in LiMn$_2$O$_4$, which was estimated to be $D_{chem} = 3.5 \times 10^{-12}$ cm$^2$ s$^{-1}$ at room temperature. Moreover, this analysis showed that the apparent pseudo-capacitive performance, previously reported for LiMn$_2$O$_4$, can be directly explained by the high surface area of the porous structure of the layer. To the best of the authors' knowledge, this is the first time that the SE technique has been used for *operando* studies of battery materials, therefore opening a new avenue for understanding mass transport properties and critical interfacial phenomena occurring during battery operation. Moreover, the high spatial resolution, multi-layer combined analysis and time-resolved capabilities of the technique will definitely contribute new insights into materials and devices at the nanoscale level for improving current performances by interface engineering.

## Conflicts of interest

There are no conflicts to declare.

## Acknowledgements

The research was supported by the Generalitat de Catalunya-AGAUR (2017 SGR 1421), the CERCA Programme, and the European Regional Development Funds (ERDF, "FEDER Programa Competitivitat de Catalunya 2007–2013"). This project has received funding from the European Research Council (ERC) under the European Union's Horizon 2020 research and innovation program (ULTRASOFC, G.A. 681146 and HARVESTORE, FET-RIA-824072).

## Notes and references

1 X. Yu and A. Manthiram, Electrode–electrolyte interfaces in lithium-based batteries, *Energy Environ. Sci.*, 2018, **11**, 527–543.

2 K. Takada, T. Ohno, N. Ohta, T. Ohnishi and Y. Tanaka, Positive and Negative Aspects of Interfaces in Solid-State Batteries, *ACS Energy Lett.*, 2018, **3**, 98–103.

3 S. Takai, *et al.* Tracer diffusion coefficients of lithium ion in LiMn2O 4 measured by neutron radiography, *Solid State Ionics*, 2014, **256**, 93–96.

4 C. Ma, *et al.* Atomic-scale origin of the large grain-boundary resistance in perovskite Li-ion-conducting solid electrolytes, *Energy Environ. Sci.*, 2014, **7**, 1638–1642.

5 S. Yang, B. Yan, L. Lu and K. Zeng, Grain boundary effects on Li-ion diffusion in a Li$_{1.2}$Co$_{0.13}$Ni$_{0.13}$Mn$_{0.54}$O$_2$ thin film cathode studied by scanning probe microscopy techniques, *RSC Adv.*, 2016, **6**, 94000–94009.

6 J. F. Wu and X. Guo, Origin of the low grain boundary conductivity in lithium ion conducting perovskites: Li$_{3x}$La$_{0.67-x}$TiO$_3$, *Phys. Chem. Chem. Phys.*, 2017, **19**, 5880–5887.

7 Y. F. Zhao, B. Lu and J. Zhang, Lithium Diffusion and Stress in a Polycrystalline Film Electrode, *Acta Mech. Solida Sin.*, 2018, **31**, 290–309.

8 P. Yan, *et al.* Tailoring grain boundary structures and chemistry of Ni-rich layered cathodes for enhanced cycle stability of lithium-ion batteries, *Nat. Energy*, 2018, **3**, 600–605.

9 J. Woods, N. Bhattarai, P. Chapagain, Y. Yang and S. Neupane, In situ transmission electron microscopy observations of rechargeable lithium ion batteries, *Nano Energy*, 2019, **56**, 619–640.

10 U. Boesenberg and U. E. A. Fittschen, 2D and 3D Imaging of Li-Ion Battery Materials Using Synchrotron Radiation Sources, in *Rechargeable Batteries: Materials, Technologies and New Trends*, ed. Z. Zhang and S. S. Zhang, Springer International Publishing, 2015, pp. 393–418, DOI: 10.1007/978-3-319-15458-9_14.

11 F. Lin, *et al.* Synchrotron X-ray Analytical Techniques for Studying Materials Electrochemistry in Rechargeable Batteries, *Chem. Rev.*, 2017, **117**, 13123–13186.

12 A. M. Tripathi, W.-N. Su and B. J. Hwang, In situ analytical techniques for battery interface analysis, *Chem. Soc. Rev.*, 2018, **47**, 736–851.

13 P. P. R. M. L. Harks, F. M. Mulder and P. H. L. Notten, In situ methods for Li-ion battery research: A review of recent developments, *J. Power Sources*, 2015, **288**, 92–105.

14 Z. Zhu, *et al.* In Situ Mass Spectrometric Determination of Molecular Structural Evolution at the Solid Electrolyte Interphase in Lithium-Ion Batteries, *Nano Lett.*, 2015, **15**, 6170–6176.

15 S. Mukerjee, *et al.* Structural Evolution of $Li_x Mn_2O_4$ in Lithium-Ion Battery Cells Measured In Situ Using Synchrotron X-Ray Diffraction Techniques, *J. Electrochem. Soc.*, 1998, **145**, 466–472.

16 C. Bressler and M. Chergui, Ultrafast X-ray Absorption Spectroscopy, *Chem. Rev.*, 2004, **104**, 1781–1812.

17 Y.-C. Chen, *et al.* In-situ synchrotron X-ray absorption studies of $LiMn_{0.25}Fe_{0.75}PO_4$ as a cathode material for lithium ion batteries, *Solid State Ionics*, 2009, **180**, 1215–1219.

18 R. Robert, *et al.* Scanning X-ray Fluorescence Imaging Study of Lithium Insertion into Copper Based Oxysulfides for Li-Ion Batteries, *Chem. Mater.*, 2012, **24**, 2684–2691.

19 D. P. Finegan, *et al.* In-operando high-speed tomography of lithium-ion batteries during thermal runaway, *Nat. Commun.*, 2015, **6**, 6924.

20 Y. Ding, Z.-F. Li, E. V. Timofeeva and C. U. Segre, In Situ EXAFS-Derived Mechanism of Highly Reversible Tin Phosphide/Graphite Composite Anode for Li-Ion Batteries, *Adv. Energy Mater.*, 2018, **8**, 1702134.

21 P. Shearing, Y. Wu, S. J. Harris and N. Brandon, In Situ X-Ray Spectroscopy and Imaging of Battery Materials, *Electrochem. Soc. Interface*, 2011, **20**, 43–47.

22 I. Buchberger, *et al.* Aging Analysis of Graphite/$LiNi_{1/3}Mn_{1/3}Co_{1/3}O_2$ Cells Using XRD, PGAA, and AC Impedance, *J. Electrochem. Soc.*, 2015, **162**, A2737–A2746.

23 X. Q. Yang, *et al.* In Situ Synchrotron X-Ray Diffraction Studies of the Phase Transitions in $Li_xMn_2O_4$ Cathode Materials, *Electrochem. Solid-State Lett.*, 1999, **2**, 157–160.

24 B. Breitung, P. Baumann, H. Sommer, J. Janek and T. Brezesinski, In situ and operando atomic force microscopy of high-capacity nano-silicon based electrodes for lithium-ion batteries, *Nanoscale*, 2016, **8**, 14048–14056.

25 N. E. Drewett, I. M. Aldous, J. Zou and L. J. Hardwick, In situ Raman spectroscopic analysis of the lithiation and sodiation of antimony microparticles, *Electrochim. Acta*, 2017, **247**, 296–305.

26 A. Krause, *et al.* In Situ Raman Spectroscopy on Silicon Nanowire Anodes Integrated in Lithium Ion Batteries, *J. Electrochem. Soc.*, 2019, **166**, A5378–A5385.

27 J. Lei, F. McLarnon and R. Kostecki, In Situ Raman Microscopy of Individual $LiNi_{0.8}Co_{0.15}Al_{0.05}O_2$ Particles in a Li-Ion Battery Composite Cathode, *J. Phys. Chem. B*, 2005, **109**, 952–957.

28 M. Matsui, H. Kuwata and N. Imanishi, Operando FTIR Spectroscopy for Lithium-Ion Batteries, *Meet. Abstr.*, 2014, 247.

29 D. Streich and P. Novák, Electrode-electrolyte interface characterization of carbon electrodes in $Li-O_2$ batteries: capabilities and limitations of infrared spectroscopy, *Electrochim. Acta*, 2016, **190**, 753–757.

30 J. Lei, L. Li, R. Kostecki, R. Muller and F. McLarnon, Characterization of SEI Layers on $LiMn_2O_4$ Cathodes with In Situ Spectroscopic Ellipsometry, *J. Electrochem. Soc.*, 2005, **152**, A774–A777.

31 H.-G. Tompkins and E.-A. Irene, *Handbook of Ellipsometry*, ed. H. G. Tompkins and E. A. Irene, Springer, Berlin, 2005, 860, p. 3-540-22293-6.

32 M. M. Thackeray, W. I. F. David, P. G. Bruce and J. B. Goodenough, Lithium insertion into manganese spinels, *Mater. Res. Bull.*, 1983, **18**, 461–472.

33 T. Ohzuku and R. J. Brodd, An overview of positive-electrode materials for advanced lithium-ion batteries, *J. Power Sources*, 2007, **174**, 449–456.

34 F. Jiao, J. Bao, A. H. Hill and P. G. Bruce, Synthesis of Ordered Mesoporous Li–Mn–O Spinel as a Positive Electrode for Rechargeable Lithium Batteries, *Angew. Chem., Int. Ed.*, 2008, **47**, 9711–9716.

35 M. Okubo, *et al.* Fast Li-Ion Insertion into Nanosized $LiMn_2O_4$ without Domain Boundaries, *ACS Nano*, 2010, **4**, 741–752.

36 W. Tang, *et al.* An aqueous rechargeable lithium battery of excellent rate capability based on a nanocomposite of $MoO_3$ coated with PPy and $LiMn_2O_4$, *Energy Environ. Sci.*, 2012, **5**, 6909–6913.

37 W. Tang, *et al.* $LiMn_2O_4$ Nanotube as Cathode Material of Second-Level Charge Capability for Aqueous Rechargeable Batteries, *Nano Lett.*, 2013, **13**, 2036–2040.

38 Z. Quan, S. Ohguchi, M. Kawase, H. Tanimura and N. Sonoyama, Preparation of nanocrystalline $LiMn_2O_4$ thin film by electrodeposition method and its electrochemical performance for lithium battery, *J. Power Sources*, 2013, **244**, 375–381.

39 M. J. Young, H.-D. Schnabel, A. M. Holder, S. M. George and C. B. Musgrave, Band Diagram and Rate Analysis of Thin Film Spinel $LiMn_2O_4$ Formed by Electrochemical Conversion of ALD-Grown MnO, *Adv. Funct. Mater.*, 2016, **26**, 7895–7907.

40 B. K. Lesel, J. S. Ko, B. Dunn and S. H. Tolbert, Mesoporous $Li_xMn_2O_4$ Thin Film Cathodes for Lithium-Ion Pseudocapacitors, *ACS Nano*, 2016, **10**, 7572–7581.

41 B. Put, P. Vereecken, N. M. Labyedh, A. Sepulveda, C. Huyghebaert, I. Radu and A. Stesmans, High Cycling Stability and Extreme Rate Performance in Nanoscaled $LiMn_2O_4$ Thin Films, *ACS Appl. Mater. Interfaces*, 2015, **7**, 22413–22420.

42 M. Fehse, *et al.* Ultrafast dischargeable $LiMn_2O_4$ thin film electrodes with pseudocapacitive properties for microbatteries, *ACS Appl. Mater. Interfaces*, 2017, **9**, 6b15258.

43 R. Trócoli, *et al.* High Specific Power Dual-Metal-Ion Rechargeable Microbatteries Based on $LiMn_2O_4$ and Zinc for Miniaturized Applications, *ACS Appl. Mater. Interfaces*, 2017, **9**.

44 M. Fehse, *et al.* An innovative multi-layer pulsed laser deposition approach for $LiMn_2O_4$ thin film cathodes, *Thin Solid Films*, 2018, **648**, 108–112.

45 M. Stchakovsky, Y. Battie and A. En Naciri, An original method to determine complex refractive index of liquids by spectroscopic ellipsometry and illustrated applications, *Thin Solid Films*, 2017, **421**, 802–806.

46 G. E. Jellison and F. A. Modine, Parameterization of the optical functions of amorphous materials in the interband region, *Appl. Phys. Lett.*, 1996, **69**, 371–373.

47 S. Gong and B.-G. Liu, Electronic energy gaps and optical properties of $LaMnO_3$, *Phys. Lett. A*, 2011, **375**, 1477–1480.

48 K. J. Kim and J. H. Lee, Evolution of the structural and the optical properties and the related electronic structure of $LiT_xMn_{2-x}O_4$ (T = Fe and Ni) thin films, *J. Korean Phys. Soc.*, 2007, **51**, 1166–1171.

49 L. V. Nomerovannaya, A. A. Makhnëv and A. Y. Rumyantsev, Evolution of the optical properties of single-crystal $La_{1-x}Sr_xMnO_3$, *Phys. Solid State*, 1999, **41**, 1322–1326.

50 N. Kuwata, M. Nakane, T. Miyazaki, K. Mitsuishi and J. Kawamura, Lithium diffusion coefficient in $LiMn_2O_4$ thin films measured by secondary ion mass spectrometry with ion-exchange method, *Solid State Ionics*, 2018, **320**, 266–271.

51 D. Shu, K. Y. Chung, W. I. Cho and K.-B. Kim, Electrochemical investigations on electrostatic spray deposited $LiMn_2O_4$ films, *J. Power Sources*, 2003, **114**, 253–263.

52 M. Nishizawa, *et al.* Electrochemical Studies of Spinel $LiMn_2O_4$ Films Prepared by Electrostatic Spray Deposition, *Bull. Chem. Soc. Jpn.*, 1998, **71**, 2011–2015.

53 J. Xie, *et al.* Li-ion transport kinetics in $LiMn_2O_4$ thin films prepared by radio frequency magnetron sputtering, *J. Power Sources*, 2008, **180**, 576–581.

54 S. B. Tang, M. O. Lai and L. Lu, Study on $Li^+$-ion diffusion in nano-crystalline $LiMn_2O_4$ thin film cathode grown by pulsed laser deposition using CV, EIS and PITT techniques, *Mater. Chem. Phys.*, 2008, **111**, 149–153.

55 T. Okumura, *et al.* Determination of lithium ion diffusion in lithium-manganese-oxide-spinel thin films by secondary-ion mass spectrometry, *J. Power Sources*, 2009, **189**, 643–645.

56 K. A. Striebel, C. Z. Deng, S. J. Wen and E. J. Cairns, Electrochemical Behavior of $LiMn_2O_4$ and $LiCoO_2$ Thin Films Produced with Pulsed Laser Deposition, *J. Electrochem. Soc.*, 1996, **143**, 1821–1827.

57 C. Julien, E. Haro-Poniatowski, M. A. Camacho-Lopez, L. Escobar-Alarcon and J. Jimenez-Jarquin, Growth of $LiMn_2O_4$ thin films by pulsed-laser deposition and their electrochemical properties in lithium microbatteries, *Mater. Sci. Eng., B*, 2000, **72**, 36–46.